# DYNAMIC RESOURCE MANAGEMENT IN CLOUD DATA CENTERS FOR SERVER CONSOLIDATION

ALEXANDER NGENZI

*PhD Student , Department of Computer Science Engineering, Jain University, Bangalore-India*
*yngenzi37@gmail.com*

Dr. SELVARANI R

*Professor ,Computer Science Engineering, Alliance University, Bangalore, India*
*selvarani.riic@gmail.com*

Dr. SUCHITHRA R NAIR

*HoD, Department of Master of Science in Information Technology, Jain University, Bangalore . India.*
*suchithra.suriya@gmail.com*

**ABSTRACT:** Cloud resource management has been a key factor for the cloud datacenters development. Many cloud datacenters have problems in understanding and implementing the techniques to manage, allocate and migrate the resources in their premises. The consequences of improper resource management may result into underutilized and wastage of resources which may also result into poor service delivery in these datacenters. Resources like; CPU, memory, Hard disk and servers need to be well identified and managed. In this Paper, Dynamic Resource Management Algorithm(DRMA) shall limit itself in the management of CPU and memory as the resources in cloud datacenters. The target is to save those resources which may be underutilized at a particular period of time. It can be achieved through Implementation of suitable algorithms. Here, Bin packing algorithm can be used whereby the best fit algorithm is deployed to obtain results and compared to select suitable algorithm for efficient use of resources.

**KEYWORDS-** Best-fit Bin packing ,Cloud computing, CPU , DRMA ,Server consolidation.

## I. INTRODUCTION

Large Cloud datacenters comprise of many thousands of servers and most of the time these servers are underutilized. Cloud computing is a technology that uses the Internet and central remote servers to maintain data and applications. Cloud computing that has become an increasingly important trend, is a virtualization technology that uses the internet and central remote servers to offer the sharing of resources that include infrastructures, software, applications and business processes to the market environment to fulfill the elastic demand. The massive amount of wastage of resources in Cloud datacenters results in resource management problems. The challenges related to datacenters with a particular emphasis on how new virtualization technologies can be used to simplify deployment, improve resource efficiency and reduce the number of usage of physical servers can be taken for study. In this paper, Cloud Resource Management algorithm will be discussed where by suitable algorithms will be deployed for the proper resource allocations and migrations to achieve efficient resource management. The term resource management refers to the operations used to control how capabilities provided by Cloud resources and services are made available to other entities, whether users, applications, services. In this paper, two configurable resources have been identified which are CPU and Memory. There are many cloud data resources but this research will limit itself on the mentioned resources. This is mainly due to the fact that the resources will help us to achieve the fast execution and accessibility as well as storage facilities. Central Processing Unit (CPU) is expected to provide fast execution as far as resources in datacenters are concerned and memory will provide storage facilities, accessibility and availability in cloud datacenters. Bin packing algorithm can be used. Bin Packing Problem is a problem where objects with a given area must be packed into a finite number of bins with a given area such that the minimum amount of bins are used. Here the objects to be packed are the servers being consolidated and the bins are the targeted servers.

## II. RELATED WORK

Cloud datacenters need efficient resource management that is able to orchestrate bulks of different resources. In current public cloud datacenters, there is a mismatch between the requirements of tenant requests and resource utilization. Multiple resource types in datacenters make the situation even more complex[1]. As virtual machines dynamically enter and leave a cloud system, it becomes necessary to relocate virtual machines among servers. However, relocation of virtual machines introduces run-time overheads and consumes extra energy, thus an careful planning for relocation is necessary. A virtual machines can be deployed in any physical server that supports virtualization, therefore the deployment becomes very flexible, and easy to manage[2]. Due to the highly dynamic heterogeneity of resources on cloud computing platform, virtual machines must adapt to the cloud computing environment dynamically so as to achieve its best performance by fully using its service and resources. But in order to improve resource utility, resources must be properly allocated and load balancing must be guaranteed[3]. Consolidating multiple underutilized servers into a fewer number of non-dedicated servers that can host multiple applications is an effective tool for businesses to enhance their returns on investment. The problem can be modelled as a variant of the bin packing problem where items to be packed are the servers being consolidated and bins are the target servers[4]. Resources like CPU, Memory and Hard disk need to be kept minimum to avoid un necessary costs. Autonomic resource management could lead to efficient resource utilization and fast response in the presence changing workloads. The objective is to switch on those servers which are required and turn off those not required. Live migration of VMs is an essential tool for the management of Cloud Computing resources[5]. Cloud computing resource covers all useful entities which can be used through the cloud platform, including computer software, computer hardware, equipment, instrument and so on. Cloud computing resources is decided by the characteristics of cloud computing resource management system, which should have functions and features like: hiding the heterogeneity of cloud computing resource, providing users with the unified access interface, Shielding the dynamic of cloud computing resources, evaluating and estimating the performance of each resource, guaranteeing to meet the service quality of the user request ( QoS ),  a careful review of the user's request of cloud computing and ensuring the security of cloud computing[6]. As the prevalence of Cloud computing continues to grow, the need for resource management within the infrastructure layer also increases. The provisioning of the Cloud infrastructure in the data center (DC) is a fundamental prerequisite. This is followed by how well resources are allocated, migrated and managed. Also, dynamic usage patterns of the user, location and geographical distribution of DC, availability of Internet and Cloud service adaptation are all key factors in Cloud resource management[7].

As the prevalence of Cloud computing continues to grow, the need for resource management within the infrastructure layer also increases. The provisioning of the Cloud infrastructure in the data center (DC) is a fundamental prerequisite. This is followed by how well resources are allocated, migrated and managed. Also, dynamic usage patterns of the user, location and geographical distribution of DC, availability of Internet and Cloud service adaptation are all key factors in Cloud resource management. Resource management is one of the main services that both providers and customers must ensure[8]. The minimization of resources, both in CPU power and memory usage, brings benefits for both actors. Minimizing the amount of hardware resource and power consumption in use is one of the main services that such a cloud infrastructure must ensure. This objective can be done either by the customer at the application level (by dynamically sizing the application based on the workload), or by the provider at the virtualization level (by consolidating virtual machines based on the infrastructure's utilization rate)[22]. There are many ways to provide resource management controls in a cloud environment. VMware's Distributed Resource Scheduler (DRS) for  example has a rich set of controls providing the services needed for successful multi-resource management while providing differentiated QoS to groups of VMs, albeit at a small scale compared to typical cloud deployments[23].The modelling of the relocation problem as a modified bin packing problem and proposing a new server consolidation algorithm that guarantees server consolidation with bounded relocation costs can also be taken into consideration. The conduction of a detailed analysis on the complexity of the server consolidation problem, and giving an upper bound on the cost of relocation is also important. Finally, the conduction of simulations and comparison of server consolidation algorithm with other relocation methods, like First Fit and Best Fit method[2]. The characteristics of bin packing application is the necessity to pack or fit a collection of objects into a well defined regions so that they do not overlap. From engineering point of view the problem is normally one of making efficient use of time or space[24].

### III. SERVER CONSOLIDATION MODELLING

Several constraints have been taken into considerations: First, we have considered the issues related to live migration like re-allocation of tasks, migrations and so on. The live migration requires compatible utilization of simulation software like CloudSim through Java programming. This will help us to come up with configurable resources like CPU and Memory. Second, it is necessary to limit the upper bound of our resources not reaching 100% utilization threshold. This is  to prevent performance degradation. Again, live migration technology consumes high CPU cycles and high server power which would result into low performance and low throughput.

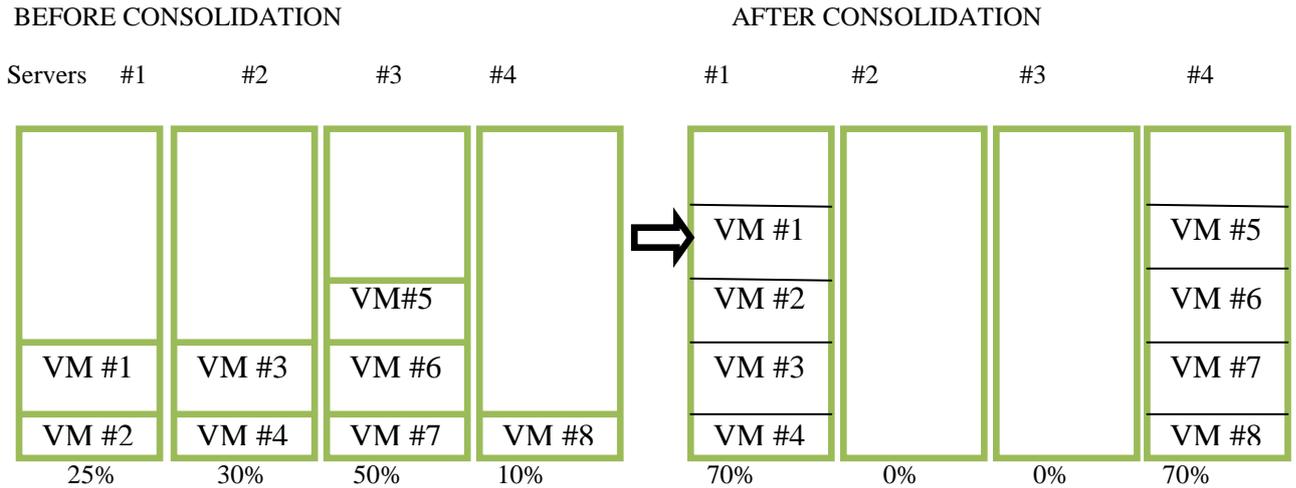

**Figure 1: Server consolidation modelling**

So, keeping CPU utilization below threshold value would result into a certain level of throughput. Choosing the right value for CPU or memory is important, as a very high threshold means that the performance of tasks running on a particular server may drop significantly, while very low threshold lowers effectiveness of consolidation. There is no fixed amount of how much percentages should be used about the optimal CPU or memory threshold values for consolidation. Hence, our experiment will take the range between 50% to 80%. Furthermore, there is a need to migrate the task from the server only if this results in releasing this server. Server consolidation modelling(Figure 1) has some properties of Bin packing algorithm, First-First and Best-Fit. However, First-Fit and Best-Fit are targeted only at minimizing number of servers while server consolidation modelling is targeted minimizing the number of migrations as well. As it is seen, after migration servers#2 and #3 are turned off to minimize power consumption as well as resources which may be wasted.

This is aimed at minimizing the number of migrations, but in the case of migrating without releasing the server, it does not help us to achieve our objective. Rather, the first thing is to determine which tasks have to be moved to which servers, and after execution of algorithm then migrations are accomplished as per the results attained. In this research work, two dimensions can be used to characterize the task and server. Let the number of servers be *n* and the number of tasks be *m*.

Suppose $t_{ij}$ is the memory required for $i^{th}$ server and $j^{th}$ job(task) and $t_s$ is the free space for i= 1,2,......n and j= 1,2..........m then,

$T_{ij}$= max{ $t_{si}$ , $t_{sj}$ } i.e for i>j i=1,2,........n and j=1,2,.........m     (1)

If $T_{ij} \geq t_{ij}$ , then we go for migration     (2)

Suppose also $M_{ij}$ is the migration quantity(memory) then,

$M_{ij}$ = min{ $t_{si}$ , $t_{sj}$ }  i.e for j>i     (3)

If $M_{ij} > t_{ij}$, then it is not possible to allocate the current task.     (4)

After applying server consolidation modelling, the state of the server will change. Three metrics can be estimated with the effectiveness of server consolidation operations:

1. Number of servers used
2. Number of servers released
3. Number of tasks to be migrated.

Bin packing algorithm considers the first number as its effectiveness metrics.

Server consolidation modelling should minimize the first and third numbers. If the target is the first number then a better result can be achieved by re allocating many tasks. This is what the bin packing problems do. However, this is not desirable due to the cost of live migration described earlier. Hence, the target possibly is the optimal result that considers the third number.

## IV. PROPOSED ALGORITHM: DRMA

The Dynamic Resource Management Algorithm(DRMA) has three major components: An allocation algorithm, Migration algorithm and combined algorithm of both allocation and migration of tasks onto the server. In this section, we briefly describe these components so as the reader will have a global picture of DRMA before delving into the details.

The algorithm uses the best-fit concept to minimize the server cost. The algorithm iterates over a list of tasks sorted based on the CPU requirement in the descending order. So in the first iteration the algorithm gets the task with highest CPU requirement. Let's call the task in hand as current task. Once current task is decided, the algorithm tries finding the server on to which the task can be fitted exactly, means the server with exactly equal free CPU available as of the CPU requirement of the current task. If the algorithm fails to find the exact match, then it goes on to find the closest match, meaning that the algorithm will pick a server from the sorted server list that has the closest free CPU available with respect to CPU requirement of the current task. Once a closest match is found, the algorithm allocates the current task on to server. The algorithm continues by choosing the next task from the sorted task list. The algorithm starts again with finding out the exact match followed by finding out the closest match. Once there's no server that can accommodate the current task, the algorithm hands over the control to migration algorithm, which migrates the tasks across the servers to consolidate the running tasks.

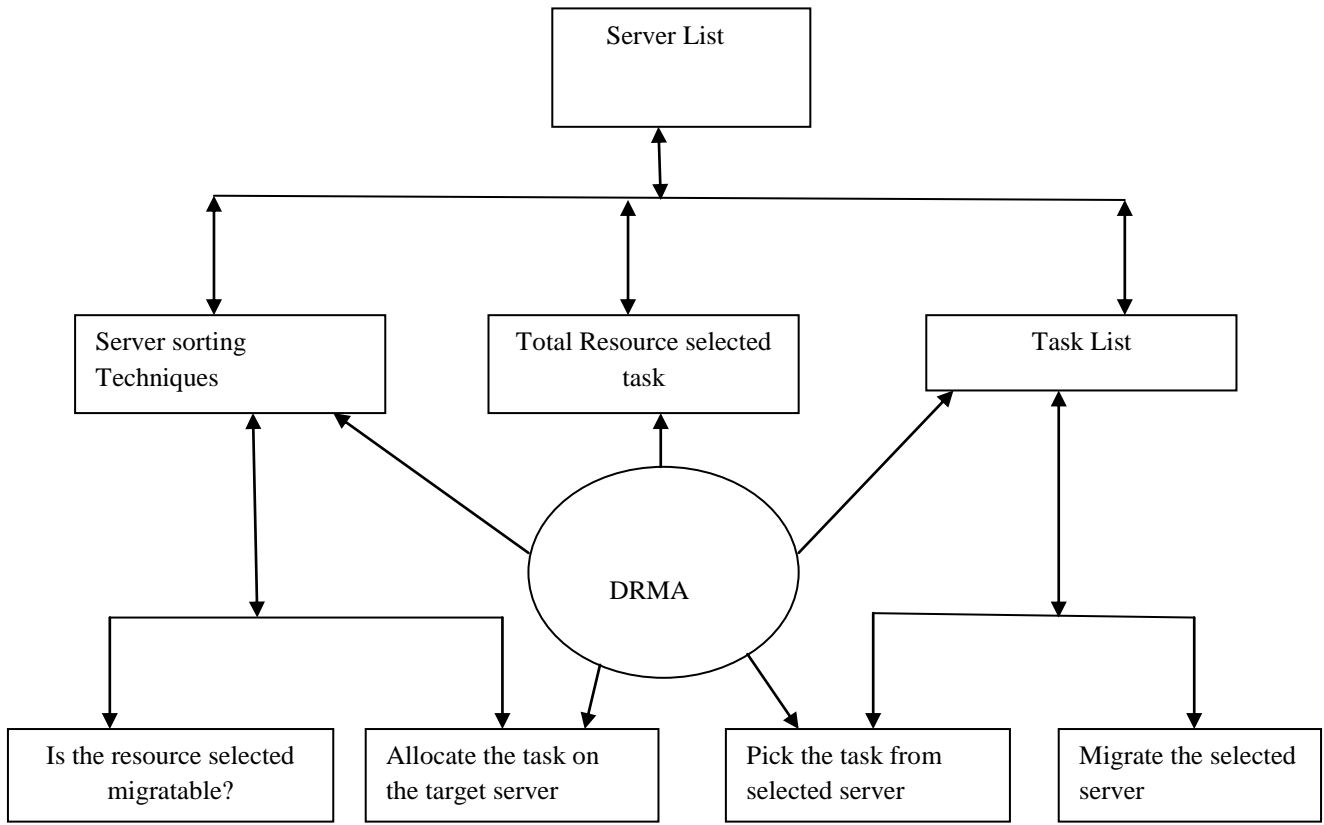

**Figure 2: Framework of the Proposed algorithm(DRMA)**

As the algorithm is choosing the best possible server for allocation, it prevents the wastage of free CPU available on the servers.

The algorithm iterates over a list of servers sorted based free CPU available in the ascending order. Initially it chooses the first server from the sorted server list i.e., the server with least free CPU available or in other words, the server which has utilized the most of its CPU. Let's call the selected server as the target server. Once the target server is selected, the algorithm makes a list of running tasks on all servers leaving the tasks from the target server. Then algorithm sorts this list in ascending order based on the CPU requirement of the tasks. A task from this sorted task list is chosen as the task to be migrated. Once the task is chosen, the algorithm checks whether the task can be migrated to target server? If it can be migrated, then it checks whether the migration results in making enough free CPU available to allocate the current task from waiting task list? If the migration results in making enough free CPU available for allocation of current task, then the actual migration is made i.e., the task selected from sorted task list is moved to the target server and the current task allocated on to the server from where the task is moved. If the migration of selected task is not possible or if it's yielding enough free CPU for allocation of current task, then the algorithm goes on to choose the next from the sorted task list to try with migration. Like this the algorithm iterates over all the tasks from the sorted task list to find a feasible migration that results in allocation of the current task. If no feasible migration is possible, then it's concluded that there's no way to allocate the current task.

**DRM ALGORITHM**

*Input*: C, S, T
Where
    C <- Configurable Parameters like memory, CPU
    $C_k$ <- $k^{th}$ Parameter in C
    k <- 0
    S <- List of all servers
    T <- Task List
    $T_i$ <- $i^{th}$ Task in T
BEGIN
    allocated <- false
    sorted serverlist <- sort S by $C_k$ descending
    target <- Find the server with maximum CPU available
        i.e., the first server in sorted serverlist
    tasklist <- List all the tasks except from target
    sorted tasklist <- sort T By $C_k$ descending
    foreach selectedtask belongs to sorted tasklist
        if (total available $C_k$ > $C_k$ of selectedtask)
            then   if($C_k$ of selectedtask = $C_k$ of selectedtask+1)
                then   if($C_{k+1}$ of selectedtask > $C_{k+1}$ of selectedtask+1)
                    then selectedtask <- selectedtask
                    else selectedtask <- selectedtask+1
                end if
            end if
        then foreach task belongs to target
        if (task is migratable and $C_k$ on target > $C_k$ of selectedtask)
        then migrate the task
            allocate the selectedtask on to target
            allocated <- TRUE
            break (foreach loop)
        end if
        end if
    end foreach
END

## V. Experimental setup, Environment and Results

In this experiment, we consider four servers in which tasks may be allocated or migrated. Here, Bin packing algorithm can be used to come up with configurable resources like CPU and Memory. It is necessary to limit the upper bound of our resources not reaching 100% utilization threshold. This is to prevent performance degradation. As it is explained in section IV, there is no fixed amount of how much percentages should be used about the optimal CPU or memory threshold values for consolidation. Hence, for our experiment we take the range between 40% to 70% before migration and between 70% to 100% after migration. Total utilization values have be chosen randomly for example; 70% and 40%, .This can be explained in details in the figure(3) below.

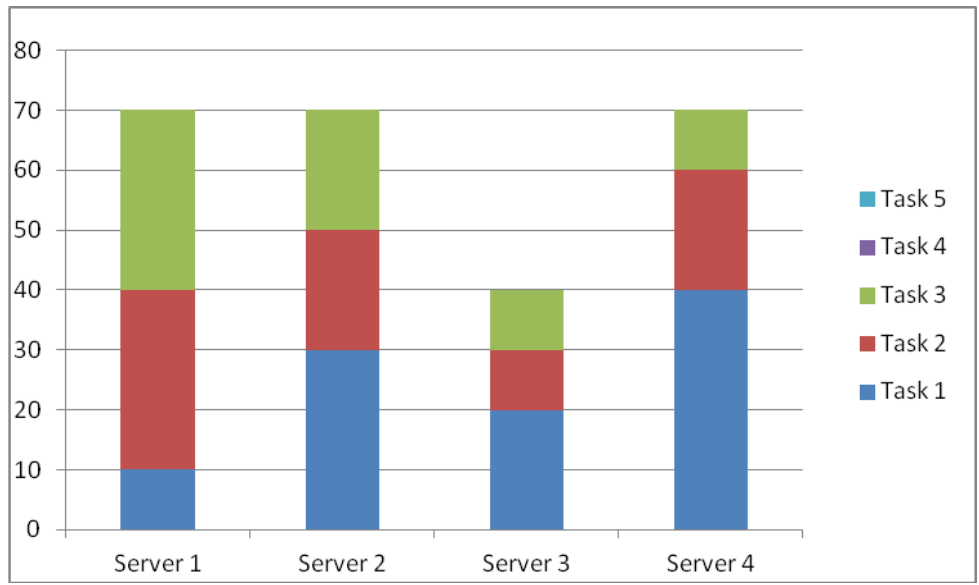

**Figure 3: Graphical presentation before migration**

**Table 1: Before Migration**

| List of Servers | Task 1 | Task 2 | Task 3 | Task 4 | Task 5 | Total |
|---|---|---|---|---|---|---|
| Server 1 | 10 | 30 | 30 | 0 | 0 | 70 |
| Server 2 | 30 | 20 | 20 | 0 | 0 | 70 |
| Server 3 | 20 | 10 | 10 | 0 | 0 | 40 |
| Server 4 | 40 | 20 | 10 | 0 | 0 | 70 |

Experiment environment is shown in Table 1 above. We implemented simulation software using Java NetBeans programming . This software basically performs operations including generating random values according to total utilization in experiment setup and analysing data, by displaying tables, graphs or statistics. The snapshots of the simulation software can be seen in both Figure 3 and Figure 4. In this software, we can specify the number of tasks and the experiment values described in section V, list of tasks and servers can also be generated in both Table 1 and Table 2.

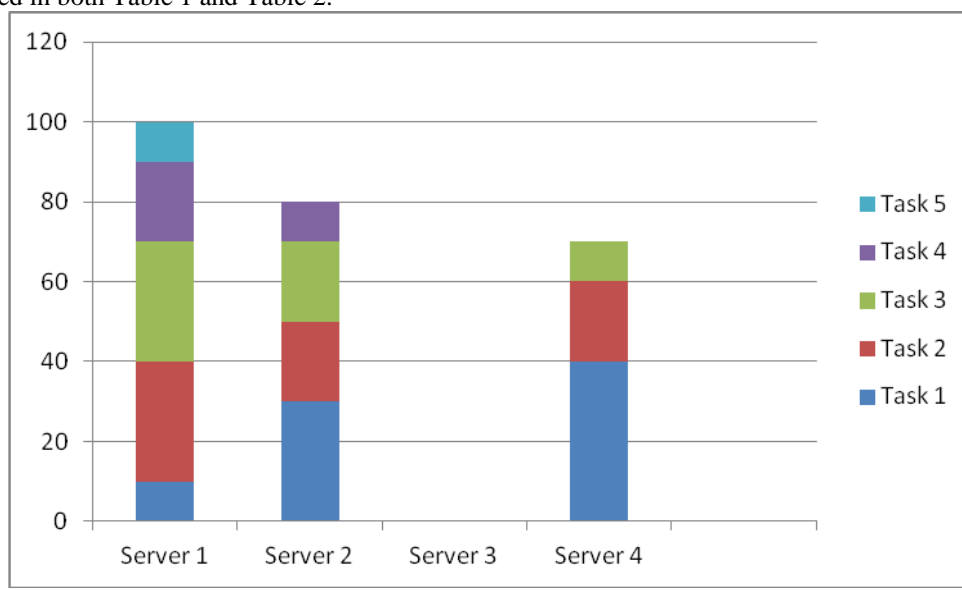

**Figure 4: Graphical presentation after migration**

**Table 2: After Migration**

| List of servers | Task 1 | Task 2 | Task 3 | Task 4 | Task 5 | Total |
|---|---|---|---|---|---|---|
| Server 1 | 10 | 30 | 30 | 20 | 10 | 100 |
| Server 2 | 30 | 20 | 20 | 10 | 0 | 80 |
| Server 3 | 0 | 0 | 0 | 0 | 0 | 0 |
| Server 4 | 40 | 20 | 10 | 0 | 0 | 70 |

As it is shown in figure 4 above, the resulting line graph shows allocation after migration.

It is showing that server 3 is not utilized and we are able to manage with the three servers though there is a fourth server available. Our objective is to use less servers with efficient migration as we are doing migration we are able to fit the tasks in three servers without using the fourth. when there is no free space in servers we migrate tasks thereby allocating resources efficiently without turning on new servers.

## VI. CONCLUSION

In this paper, we have implemented both migration and allocation algorithms for resource management in cloud datacenters using two resources, that is; CPU and memory. Bin packing algorithm has been used whereby the best fit algorithm was deployed to obtain results and compared to select suitable algorithm for efficient use of resources. The purpose of this research was to minimize underutilized and avoid over utilized resources that may perhaps create unnecessary migrations due to unpredictable workloads. The results have shown how efficiently and effectively resources could be managed, allocated and migrated among the servers. The target was to have less number of tasks/jobs migrated with maximum resource allocation, less number of servers used and amount of resources saved with minimum costs. As migrations consume additional energy and have a negative impact on the performance, before initiating a migration, the reallocation controller had to ensure that the cost of migration does not exceed the benefit.